\documentstyle[10pt,epsfig,dp_delphititle,float,hangcaption,xspace,amssymb,
amsfonts,amsmath,amsthm,cite,graphicx]{dp_delphi}
\makeindex
\def\DpPaperGroup{PH-EP}
\def\DpPaperRef{2008--012}
\def\DpDate{17 July 2008}
\def\DpAuthors{DELPHI Collaboration}
\def\DpSubmit{(Accepted by Eur. Phys. J. C)}
\def\DpTitle{{ Di-jet production in $\gamma\gamma$ collisions at LEP2}}
\def\DpComment{}
\def\DpEMail{}

\newcommand {\kt} {$k_\perp$-cluster algorithm\hspace*{1mm}}
\newcommand {\gaga} {$\gamma\gamma$ \hspace*{1mm}}
\newcommand {\ptmean} {$\overline{p_T}$ \hspace*{1mm}}

\newcommand {\agg} {$\gamma\gamma$ \hspace*{1mm}}
\newcommand {\pt} {$p_T$}
\newcommand {\xpl} {$x^+_\gamma$ \hspace*{1mm}}
\newcommand {\xmi} {$x^-_\gamma$ \hspace*{1mm}}
\newcommand {\xplmi} {($x^+_\gamma , x^-_\gamma$)}

\begin{document}
\makeatletter
\newcount\@tempcntc
\def\@citex[#1]#2{\if@filesw\immediate\write\@auxout{\string\citation{#2}}\fi
  \@tempcnta\z@\@tempcntb\m@ne\def\@citea{}\@cite{\@for\@citeb:=#2\do
    {\@ifundefined
       {b@\@citeb}{\@citeo\@tempcntb\m@ne\@citea\def\@citea{,}{\bf ?}\@warning
       {Citation `\@citeb' on page \thepage \space undefined}}%
    {\setbox\z@\hbox{\global\@tempcntc0\csname b@\@citeb\endcsname\relax}%
     \ifnum\@tempcntc=\z@ \@citeo\@tempcntb\m@ne
       \@citea\def\@citea{,}\hbox{\csname b@\@citeb\endcsname}%
     \else
      \advance\@tempcntb\@ne
      \ifnum\@tempcntb=\@tempcntc
      \else\advance\@tempcntb\m@ne\@citeo
      \@tempcnta\@tempcntc\@tempcntb\@tempcntc\fi\fi}}\@citeo}{#1}}
\def\@citeo{\ifnum\@tempcnta>\@tempcntb\else\@citea\def\@citea{,}%
  \ifnum\@tempcnta=\@tempcntb\the\@tempcnta\else
   {\advance\@tempcnta\@ne\ifnum\@tempcnta=\@tempcntb \else \def\@citea{--}\fi
    \advance\@tempcnta\m@ne\the\@tempcnta\@citea\the\@tempcntb}\fi\fi}
 
\makeatother

\begin{titlepage}
\pagenumbering{roman}

\CERNpreprint{\DpPaperGroup}{\DpPaperRef}   
\date{{\small\DpDate}}                      
\title{\DpTitle}                            
\address{\DpAuthors}                        

\begin{shortabs}                            
\noindent
The production of two high-$p_T$ jets in the interactions of quasi-real
photons in $e^+e^-$ collisions at $\sqrt{s_{ee}}$ from 189 GeV to 209 GeV 
is studied with data corresponding to an integrated $e^+e^-$ luminosity 
of 550 pb$^{-1}$. The jets reconstructed by the \kt are defined within the
pseudo-rapidity range $-1<\eta<1$  and with jet transverse momentum, $p_T$, 
above 3 GeV/c. The differential di-jet cross-section is measured as 
a function of the mean transverse momentum \ptmean of the jets and
is compared to perturbative QCD calculations.
\end{shortabs}

\vfill

\begin{center}
\DpSubmit \ \\          
\DpComment \ \\
\DpEMail \ \\
\end{center}

\vfill
\clearpage

\headsep 10.0pt

\addtolength{\textheight}{10mm}
\addtolength{\footskip}{-5mm}
\begingroup
%
\newcommand{\DpName}[2]{\hbox{#1$^{\ref{#2}}$},\hfill}
\newcommand{\DpNameTwo}[3]{\hbox{#1$^{\ref{#2},\ref{#3}}$},\hfill}
\newcommand{\DpNameThree}[4]{\hbox{#1$^{\ref{#2},\ref{#3},\ref{#4}}$},\hfill}
\newskip\Bigfill \Bigfill = 0pt plus 1000fill
\newcommand{\DpNameLast}[2]{\hbox{#1$^{\ref{#2}}$}\hspace{\Bigfill}}

%
\footnotesize
\noindent
\DpName{J.Abdallah}{LPNHE}
\DpName{P.Abreu}{LIP}
\DpName{W.Adam}{VIENNA}
\DpName{P.Adzic}{DEMOKRITOS}
\DpName{T.Albrecht}{KARLSRUHE}
\DpName{R.Alemany-Fernandez}{CERN}
\DpName{T.Allmendinger}{KARLSRUHE}
\DpName{P.P.Allport}{LIVERPOOL}
\DpName{U.Amaldi}{MILANO2}
\DpName{N.Amapane}{TORINO}
% ...
\normalsize
\endgroup
\newpage

\titlefoot{Department of Physics and Astronomy, Iowa State
     University, Ames IA 50011-3160, USA
    \label{AMES}}
\titlefoot{Physics Department, Universiteit Antwerpen,
     Universiteitsplein 1, B-2610 Antwerpen, Belgium
    \label{ANTWERP}}
\titlefoot{IIHE, ULB-VUB,
     Pleinlaan 2, B-1050 Brussels, Belgium
    \label{BRUSSELS}}
\titlefoot{Physics Laboratory, University of Athens, Solonos Str.
     104, GR-10680 Athens, Greece
    \label{ATHENS}}
\titlefoot{Department of Physics, University of Bergen,
     All\'egaten 55, NO-5007 Bergen, Norway
    \label{BERGEN}}
\titlefoot{Dipartimento di Fisica, Universit\`a di Bologna and INFN,
     Via Irnerio 46, IT-40126 Bologna, Italy
    \label{BOLOGNA}}
\titlefoot{Centro Brasileiro de Pesquisas F\'{\i}sicas, rua Xavier Sigaud 150,
     BR-22290 Rio de Janeiro, Brazil
    \label{BRASIL-CBPF}}
\titlefoot{Inst. de F\'{\i}sica, Univ. Estadual do Rio de Janeiro,
     rua S\~{a}o Francisco Xavier 524, Rio de Janeiro, Brazil
    \label{BRASIL-IFUERJ}}
\titlefoot{Coll\`ege de France, Lab. de Physique Corpusculaire, IN2P3-CNRS,
     FR-75231 Paris Cedex 05, France
    \label{CDF}}
\titlefoot{CERN, CH-1211 Geneva 23, Switzerland
    \label{CERN}}
\titlefoot{Institut de Recherches Subatomiques, IN2P3 - CNRS/ULP - BP20,
     FR-67037 Strasbourg Cedex, France
    \label{CRN}}
\titlefoot{Now at DESY-Zeuthen, Platanenallee 6, D-15735 Zeuthen, Germany
    \label{DESY}}
\titlefoot{Institute of Nuclear Physics, N.C.S.R. Demokritos,
     P.O. Box 60228, GR-15310 Athens, Greece
    \label{DEMOKRITOS}}
\titlefoot{FZU, Inst. of Phys. of the C.A.S. High Energy Physics Division,
     Na Slovance 2, CZ-182 21, Praha 8, Czech Republic
    \label{FZU}}
\titlefoot{Dipartimento di Fisica, Universit\`a di Genova and INFN,
     Via Dodecaneso 33, IT-16146 Genova, Italy
    \label{GENOVA}}
\titlefoot{Institut des Sciences Nucl\'eaires, IN2P3-CNRS, Universit\'e
     de Grenoble 1, FR-38026 Grenoble Cedex, France
    \label{GRENOBLE}}
\titlefoot{Helsinki Institute of Physics and Department of Physical Sciences,
     P.O. Box 64, FIN-00014 University of Helsinki, 
     \indent~~Finland
    \label{HELSINKI}}
\titlefoot{Joint Institute for Nuclear Research, Dubna, Head Post
     Office, P.O. Box 79, RU-101 000 Moscow, Russian Federation
    \label{JINR}}
\titlefoot{Institut f\"ur Experimentelle Kernphysik,
     Universit\"at Karlsruhe, Postfach 6980, DE-76128 Karlsruhe,
     Germany
    \label{KARLSRUHE}}
\titlefoot{Institute of Nuclear Physics PAN,Ul. Radzikowskiego 152,
     PL-31142 Krakow, Poland
    \label{KRAKOW1}}
\titlefoot{Faculty of Physics and Nuclear Techniques, University of Mining
     and Metallurgy, PL-30055 Krakow, Poland
    \label{KRAKOW2}}
\titlefoot{LAL, Univ Paris-Sud, CNRS/IN2P3, Orsay, France
    \label{LAL}}
\titlefoot{School of Physics and Chemistry, University of Lancaster,
     Lancaster LA1 4YB, UK
    \label{LANCASTER}}
\titlefoot{LIP, IST, FCUL - Av. Elias Garcia, 14-$1^{o}$,
     PT-1000 Lisboa Codex, Portugal
    \label{LIP}}
\titlefoot{Department of Physics, University of Liverpool, P.O.
     Box 147, Liverpool L69 3BX, UK
    \label{LIVERPOOL}}
\titlefoot{Dept. of Physics and Astronomy, Kelvin Building,
     University of Glasgow, Glasgow G12 8QQ, UK
    \label{GLASGOW}}
\titlefoot{LPNHE, IN2P3-CNRS, Univ.~Paris VI et VII, Tour 33 (RdC),
     4 place Jussieu, FR-75252 Paris Cedex 05, France
    \label{LPNHE}}
\titlefoot{Department of Physics, University of Lund,
     S\"olvegatan 14, SE-223 63 Lund, Sweden
    \label{LUND}}
\titlefoot{Universit\'e Claude Bernard de Lyon, IPNL, IN2P3-CNRS,
     FR-69622 Villeurbanne Cedex, France
    \label{LYON}}
\titlefoot{Dipartimento di Fisica, Universit\`a di Milano and INFN-MILANO,
     Via Celoria 16, IT-20133 Milan, Italy
    \label{MILANO}}
\titlefoot{Dipartimento di Fisica, Univ. di Milano-Bicocca and
     INFN-MILANO, Piazza della Scienza 3, IT-20126 Milan, Italy
    \label{MILANO2}}
\titlefoot{IPNP of MFF, Charles Univ., Areal MFF,
     V Holesovickach 2, CZ-180 00, Praha 8, Czech Republic
    \label{NC}}
\titlefoot{NIKHEF, Postbus 41882, NL-1009 DB
     Amsterdam, The Netherlands
    \label{NIKHEF}}
\titlefoot{National Technical University, Physics Department,
     Zografou Campus, GR-15773 Athens, Greece
    \label{NTU-ATHENS}}
\titlefoot{Physics Department, University of Oslo, Blindern,
     NO-0316 Oslo, Norway
    \label{OSLO}}
\titlefoot{Dpto. Fisica, Univ. Oviedo, Avda. Calvo Sotelo
     s/n, ES-33007 Oviedo, Spain
    \label{OVIEDO}}
\titlefoot{Department of Physics, University of Oxford,
     Keble Road, Oxford OX1 3RH, UK
    \label{OXFORD}}
\titlefoot{Dipartimento di Fisica, Universit\`a di Padova and
     INFN, Via Marzolo 8, IT-35131 Padua, Italy
    \label{PADOVA}}
\titlefoot{Rutherford Appleton Laboratory, Chilton, Didcot
     OX11 OQX, UK
    \label{RAL}}
\titlefoot{Dipartimento di Fisica, Universit\`a di Roma II and
     INFN, Tor Vergata, IT-00173 Rome, Italy
    \label{ROMA2}}
\titlefoot{Dipartimento di Fisica, Universit\`a di Roma III and
     INFN, Via della Vasca Navale 84, IT-00146 Rome, Italy
    \label{ROMA3}}
\titlefoot{DAPNIA/Service de Physique des Particules,
     CEA-Saclay, FR-91191 Gif-sur-Yvette Cedex, France
    \label{SACLAY}}
\titlefoot{Instituto de Fisica de Cantabria (CSIC-UC), Avda.
     los Castros s/n, ES-39006 Santander, Spain
    \label{SANTANDER}}
\titlefoot{Inst. for High Energy Physics, Serpukov
     P.O. Box 35, Protvino, (Moscow Region), Russian Federation
    \label{SERPUKHOV}}
\titlefoot{J. Stefan Institute, Jamova 39, SI-1000 Ljubljana, Slovenia
    \label{SLOVENIJA1}}
\titlefoot{Laboratory for Astroparticle Physics,
     University of Nova Gorica, Kostanjeviska 16a, SI-5000 Nova Gorica, Slovenia
    \label{SLOVENIJA2}}
\titlefoot{Department of Physics, University of Ljubljana,
     SI-1000 Ljubljana, Slovenia
    \label{SLOVENIJA3}}
\titlefoot{Fysikum, Stockholm University,
     Box 6730, SE-113 85 Stockholm, Sweden
    \label{STOCKHOLM}}
\titlefoot{Dipartimento di Fisica Sperimentale, Universit\`a di
     Torino and INFN, Via P. Giuria 1, IT-10125 Turin, Italy
    \label{TORINO}}
\titlefoot{INFN,Sezione di Torino and Dipartimento di Fisica Teorica,
     Universit\`a di Torino, Via Giuria 1,
     IT-10125 Turin, Italy
    \label{TORINOTH}}
\titlefoot{Dipartimento di Fisica, Universit\`a di Trieste and
     INFN, Via A. Valerio 2, IT-34127 Trieste, Italy
    \label{TRIESTE}}
\titlefoot{Istituto di Fisica, Universit\`a di Udine and INFN,
     IT-33100 Udine, Italy
    \label{UDINE}}
\titlefoot{Univ. Federal do Rio de Janeiro, C.P. 68528
     Cidade Univ., Ilha do Fund\~ao
     BR-21945-970 Rio de Janeiro, Brazil
    \label{UFRJ}}
\titlefoot{Department of Radiation Sciences, University of
     Uppsala, P.O. Box 535, SE-751 21 Uppsala, Sweden
    \label{UPPSALA}}
\titlefoot{IFIC, Valencia-CSIC, and D.F.A.M.N., U. de Valencia,
     Avda. Dr. Moliner 50, ES-46100 Burjassot (Valencia), Spain
    \label{VALENCIA}}
\titlefoot{Institut f\"ur Hochenergiephysik, \"Osterr. Akad.
     d. Wissensch., Nikolsdorfergasse 18, AT-1050 Vienna, Austria
    \label{VIENNA}}
\titlefoot{Inst. Nuclear Studies and University of Warsaw, Ul.
     Hoza 69, PL-00681 Warsaw, Poland
    \label{WARSZAWA}}
\titlefoot{Now at University of Warwick, Coventry CV4 7AL, UK
    \label{WARWICK}}
\titlefoot{Fachbereich Physik, University of Wuppertal, Postfach
     100 127, DE-42097 Wuppertal, Germany \\
\noindent
{$^\dagger$~deceased}
    \label{WUPPERTAL}}
\addtolength{\textheight}{-10mm}
\addtolength{\footskip}{5mm}
\clearpage

\headsep 30.0pt
\end{titlepage}

%
\pagenumbering{arabic}                              
\setcounter{footnote}{0}                            %
\large
\section{Introduction}
\par This paper presents the results of a study of di-jet events produced 
in two-photon collisions in the anti-tagged mode, i.e. when both scattered 
electrons\footnote{Throughout this paper, electron stands for electron and 
positron.} escape detection. Large $p_T$ processes involving 
quasi-real photons are sensitive to both, quark and gluon, components 
of the resolved photons. Thus, the analysis of high-$p_T$ jet production 
complements the studies of the deep-inelastic structure function of a quasi-real photon 
which probe the quark distribution. Considered together, they allow the parton 
density function of the photon to be determined. The perturbative QCD scale of the 
hard interactions is provided by the jet transverse energy to be used in the
calculations. Available leading and next-to-leading order calculations \cite{theor}
can be tested with the large samples of LEP-2 data. The present analysis adds new results
to those obtained by other LEP experiments \cite{otheropal,otherl3}.
\par Gamma-gamma collisions exhibit a behavior typical of hadron-hadron
interactions, i.e. the centre-of-mass frame of the hard scattered partons is moving
in the \gaga centre-of-mass frame. The $k_\perp$-cluster algorithm \cite{ktc,cat,sey},
invariant under boosts along the beam axis, is used. The variables adopted for
the jet analysis are the transverse momentum \pt, the azimuthal angle $\phi$ and
the pseudo-rapidity $\eta=-\ln(\tan(\theta/2))$ of the particle(jet)\footnote{The 
origin of the DELPHI reference system is at the centre of the detector.
The z-axis is parallel to the $e^-$ beam,
the x-axis points horizontally to the centre of the LEP ring and the y-axis
points vertically upwards. The coordinates $R,\phi$,z form a cylindrical coordinate system
and $\theta$ is the polar angle with respect to the z-axis.}.
\par
\begin{figure}[ht!]
\begin{center}
\vspace*{1cm}
\hspace*{-2.cm}
\epsfig{file=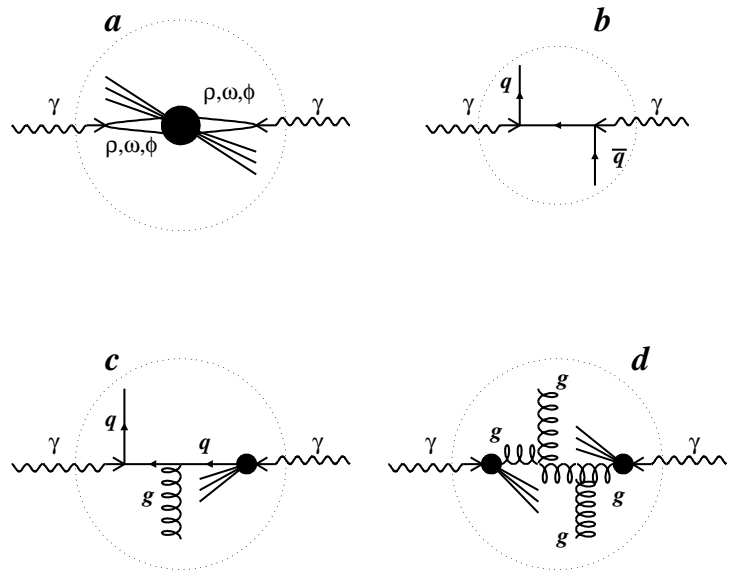,width=10cm}
\end{center}
\vspace*{-2cm}
\caption[]{Main diagrams corresponding to the hadron production in \gaga interactions
via vector meson interactions (VDM-like, a), point-like interactions (QPM-like, b)
and with one (c) or both (d) photons resolved into partons.}
\label{fig:fig1}
\end{figure}

 Hadron production in the collisions of quasi-real photons is described by
the set of processes illustrated in Fig. \ref{fig:fig1}.
The interaction of bare photons (the direct term) is described by 
the Born-box diagram within the quark-parton model (QPM). If one or both photons 
are resolved into a partonic structure,
the process is termed single- or double-resolved, respectively.
A part of the double-resolved interactions with both photons resolved 
into a bound quark-pair system is described by the vector dominance model 
(VDM). The relative contribution of the different components depends on the kinematic regime.
The present study involves such a 3-component model in order to describe the data and to correct 
on the experimental effects and then to compare the data with 
leading order (LO) and next-to-leading order (NLO) QCD calculations.
\par An important issue is the expected difference in topology.
Almost all hadrons produced in QPM-like events should belong to 
two hard-jets, while two jets in events with resolved photons 
are accompanied by remnant jet(s).
The variables sensitive to such a difference are \cite{xplmi} \\
\begin{center} $x^+_\gamma = \frac{\underset{\mathrm{jets}}{\sum}(E_{\mathrm{jet}}+p_{z,\mathrm{jet}}) }
{\underset{\mathrm{part}}{\sum}(E_{\mathrm{part}}+p_{z,\mathrm{part}})}$ 
\hspace*{0.5cm} and \hspace*{0.5cm} 
$x^-_\gamma = \frac{\underset{\mathrm{jets}}{\sum}(E_{\mathrm{jet}}-p_{z,\mathrm{jet}}) }
{\underset{\mathrm{part}}{\sum}(E_{\mathrm{part}}-p_{z,\mathrm{part}})}$,
\end{center}
where `part' corresponds to all detected particles and $E_{\mathrm{jet}}$ and $p_{z,\mathrm{jet}}$
are the two hard-jets energy and the component of jet momentum along the z-axis, respectively.
These variables represent the respective fractions of the $\gamma$ momenta
relevant to the hard
interaction. The photons in the QPM-like events participate in the interaction
entirely and both \xpl and \xmi should be equal to unity, while the presence of a
remnant jet (single-resolved photon) shifts \xpl or \xmi toward some lower values.
In the double-resolved case, both variables are far from unity.
\par Experimental details are discussed in Section 2. Section 3 presents the analysis
of the data which are compared to the leading and next-to-leading order 
QCD calculations \cite{theor}. Results and conclusions are given in Section 4.

\section{Detector and Data}
The DELPHI apparatus and performance are described in detail in ref. \cite{delphi,delper}.
\par Charged particles are detected in the tracking system
containing a 3-layer silicon Vertex Detector, the Inner Detector,
the Time Projection Chamber and the Outer Detector. The Forward Chambers 
extend the acceptance into the forward region. The tracking detectors are located inside 
the superconducting solenoid
providing a uniform magnetic field of 1.23 T parallel to the axis
of the colliding $e^+/e^-$ beams.
The combined momentum resolution provided by the tracking
system is of the order of a few per-mille in the momentum range of the present study.
\par Photons are detected in the electromagnetic calorimeters -
High density Projection Chamber in the barrel region and the Forward 
Electro-Magnetic Calorimeter in the endcaps with polar 
angle coverages of ($40^\circ - 140^\circ$) and 
($9^\circ - 35^\circ, 145^\circ - 171^\circ$), respectively.
\par Jets are reconstructed from two kinds
of detected objects - charged particles and photons.
\par The events of interest are quite energetic and the multiplicity of
final state particles is large.
The events are triggered by several components of the DELPHI trigger 
system \cite{trig} and the resulting efficiency is estimated to be equal 
to unity.
\par The data used for this analysis have been collected  by the DELPHI
detector at LEP at an $e^+e^-$ centre-of-mass energy ($\sqrt{s_{ee}}$)
from 189 GeV to 209 GeV
and correspond to an integrated $e^+e^-$ luminosity of 550 pb$^{-1}$.
\subsection{Data selection}
\par The following criteria are applied to select the data sample.
\begin{itemize}
\item {\bf Multihadron event selection.} \\
Each event should contain at least 5 charged particles.
A charged particle is counted if its momentum is greater than 0.2 GeV/c,
the polar angle is within the interval from $20^\circ$ to $160^\circ$,
the relative momentum error is less than $100\%$ and
the impact parameters are smaller than 4~cm in $R\phi$ 
and 10~cm in $z$. 
\item {\bf Background suppression.} \\
The invariant mass of the system calculated from the charged particles 
(assuming them to be pions) and the photons (detected in the electromagnetic calorimeters)
should be below 35 GeV/c$^2$. The energy threshold for an electromagnetic
calorimeter cluster is set to 0.5 GeV. The total transverse momentum $|\sum(\overrightarrow{p}_T)|$
should be below 30 GeV/c. These criteria suppress events coming from $e^+e^-$ annihilation and
they set upper limits (see below) for jet transverse momentum. 
\item {\bf Anti-tagging condition.} \\
There should be no clusters in the luminosity monitor STIC (which covers the region 
from 29 mrad to 185 mrad in the polar angle) with energy greater 
than 25 GeV (anti-tagging condition). This ensures the photons to be quasi-real.
\item {\bf Di-jet event selection.} \\
Jets are reconstructed by the $k_\perp$-cluster algorithm, implemented in the KTCLUS program \cite{sey}.
It operates with objects (at generator level they may be either particles or partons)
and tries to unite them into a jet in a recursive way.
The following variables are calculated for each object $i$ and a pair of objects
$i$ and $j$:
\begin{center} $d_i = E_{T,i}^2$ and 
$d_{ij} = min(d_i,d_j)[(\eta_i-\eta_j)^2+(\phi_i-\phi_j)^2] / R^2$, 
\end{center} where the transverse energy $E_{T,i}$ is calculated with pion mass assumption
for a charged particle and zero mass for a photon,
$R$ is usually set to unity. If the smallest value ($d_{min}$) of any $d$
in this sample is the $d_{ij}$ value of a pair $(i,j)$, then these two objects are joined
into a new object and its $E_T, \phi$ and $\eta$ are recalculated.
Otherwise, if the smallest value in the sample is the $d_i$
value of an object $i$, the corresponding object is not mergeable and
it is moved from the object list to the jet list. The procedure continues until
there are no more objects. The pseudo-rapidity $\eta$ of exactly two jets must be 
in the range $-1<\eta<1$ and their transverse momenta, \pt, above 3~GeV/c. 
Besides these two hard process jets, the event may contain one or more jets 
outside the pseudo-rapidity and $p_T$ domain mentioned above.
They are termed `remnant' jets. The choice of the kinematic limits
is conditioned by the desire to suppress soft \gaga interactions and
to keep hard process jets within the DELPHI acceptance. The cut on jet $p_T$ should not be
too large since, besides decreasing the measurement accuracy, 
it reduces the contribution coming from events with resolved photons.
The extension of the $\eta$ domain results in an increase of threshold effects, 
since some of the produced jet particles are outside the detector acceptance and 
the efficiency of the jet reconstruction is decreased.
\par An additional cut on the mean jets transverse
momentum \ptmean $> 4$ GeV/c allows a $p_T$ asymmetry of the reconstructed jets.
If such a cut is applied symmetrically to each of the two jets, the QCD calculations 
of some variables are unstable \cite{theor}.
\end{itemize}
\par The above criteria select a data sample of 5147 events.
The data are corrected for detector inefficiency and acceptance effects
with a leading-order (LO) Monte Carlo simulation described below. 
LO simulation is also used for the
estimation of the soft processes accompanying the hard-initiated jets
and for taking into account the influence of hadronisation.

\subsection{Monte Carlo simulation and background estimation}
The Monte Carlo generator PYTHIA (version 6.205) \cite{pyt} is used 
for the simulation of hadron production in \gaga interactions.
The program contains an interface to an external library of parton 
density functions (PDF) for the photons \cite{pdf}. The default SaS1D \cite{sas1d} PDF is taken.
The soft underlying events are modeled through multiple parton interactions
(MIA) of several parton pairs within the same event. An event
with both photons resolved may contain, besides hard initiated jets 
and the corresponding remnant jets, such an additional MIA contribution. 
\par The simulated sample is approximately 4 times larger than the data sample.
\par The main background process ($e^+e^- \rightarrow$ hadrons)
is simulated by the {\mbox KK2f} event generator (version 4.14) \cite{kk2f}.
The expected number of these background events in the selected data sample
is estimated to $(500\pm5)$ events.
The contamination of $\tau$ pairs produced in two-photon interactions
is evaluated as $(43\pm3)$ events, using the BDKRC program \cite{bdkrc}.
The background of $\tau$ pairs produced in $e^+e^-$ annihilation is negligible.
The events with $W, Z$ bosons contribute $(38\pm4)$ events.
\par Note that the mentioned background sources (they will be referred to as
`non-$\gamma\gamma$' background) are only a part of the total background.
Additional sources will be discussed later in Section 3.2.

\section{Analysis}
\par
\begin{figure}[ht!]
\begin{center}
\vspace*{1cm}
\hspace*{-2.cm}
\epsfig{file=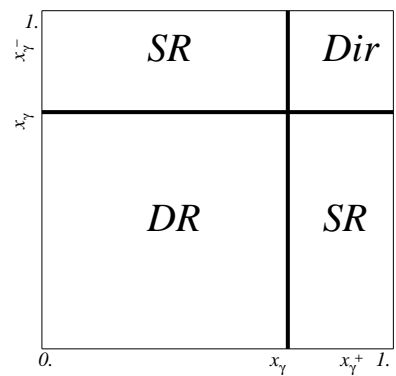,width=7cm}
\end{center}
\vspace*{-2cm}
\caption[]{\xplmi-space divided by $x_{\gamma}$ into four domains 
(`Dir', `SR' and `DR' notations are introduced in section 3.}
\label{fig:fig2}
\end{figure}

 The entire \xplmi-space can be split into four quadrants (Fig. \ref{fig:fig2}) 
by $x_{\gamma}$ equal 
to 0.85. The chosen value of $x_{\gamma}$ selects approximately equal statistics 
in the different domains. One can define three kinematic regions: the first - 
\xpl and \xmi greater than 0.85 (the region is termed as `Dir' domain), the second - 
when both values are below 0.85 (double-resolved domain, `DR') and the third region - 
when one variable is below 0.85 while the other is above it (single-resolved, `SR').

\begin{figure}[ht!]
\begin{center}
\vspace*{3cm}
\hspace*{-2cm}
\epsfig{file=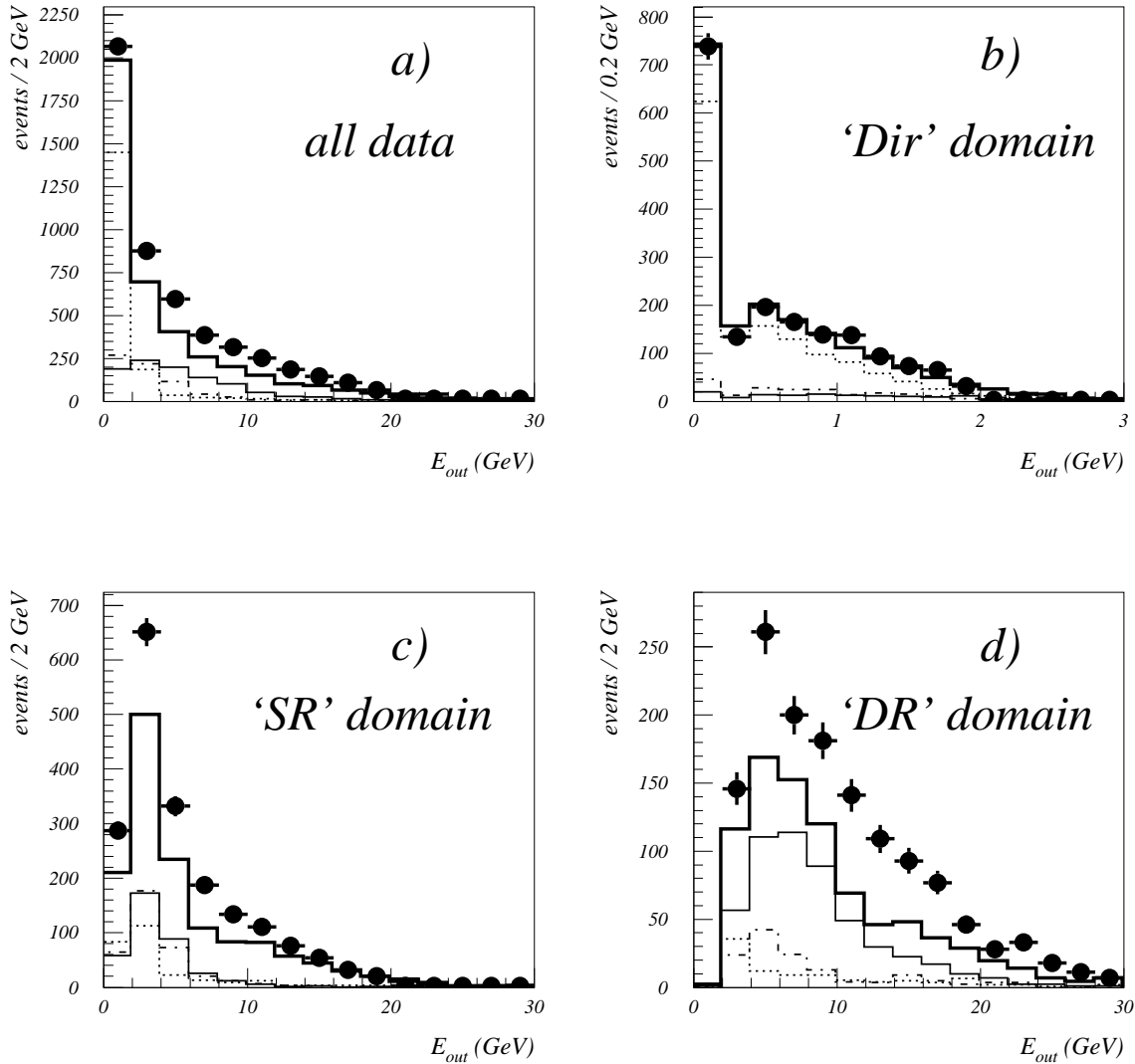,width=17cm}
\end{center}
\vspace*{-1cm}
\caption[]{Comparison of the total energy outside the
reconstructed jets $E_{out}$ with the simulation for the selected
data (a), and data in the phase-space domains `Dir' (b), `SR' (c) and `DR' (d),
described in section 3 of the text.
The thick solid histograms correspond to the sum of the di-jet \gaga interactions
as predicted by PYTHIA and the background processes.
The contributions of different \gaga subprocesses are shown by dotted (QPM term),
dashed-dotted (single-resolved) and thin solid (double-resolved) histograms.}
\label{fig:fig3}
\end{figure}

Figure \ref{fig:fig3} shows the total energy outside the 
reconstructed jets ($E_{out}$) in comparison with the model expectation.
The contributions of different parts of the model are also presented in the figure.
\par The `Dir' domain (Fig. 3b) is mostly populated by QPM-like events (94\%),
the contribution of the double-resolved subprocesses to the `DR' domain (Fig. 3d)
is around 54\%, while the `SR' domain (Fig. 3c) contains all three types in
nearly equal parts. It is seen that the model does not describe the data in the parts
of the \xplmi-space where the contribution of the resolved processes is
essential. It is also clear that a simple one-parameter renormalization 
of the model as a whole as done in previous publications \cite{otheropal,otherl3}
is not adequate since its components have to be tuned separately 
with different factors. This renormalization is a new approach to the analysis of
jets produced in $\gamma\gamma$ collisions.
\par A 3-parameters fit of the data is performed simultaneously on the 
following distributions: $E_{out}$ (Fig. 3), the transverse momentum 
balance~(~$|\sum{\overrightarrow{p}_{i,T}}|$) (not shown) and the~total invariant 
mass $W$ of the detected particles (not shown). The distributions are fitted in each 
\xplmi-domain.
The obtained parameters (scale factors $\alpha$ for each model component) are $\alpha_{qpm} = 
(0.86\pm0.02), \alpha_{s-res} = (1.49\pm0.09)$ and $\alpha_{d-res} = (1.93\pm0.05)$.
\par
\begin{figure}[ht!]
\begin{center}
\vspace*{3cm}
\hspace*{-2cm}
\epsfig{file=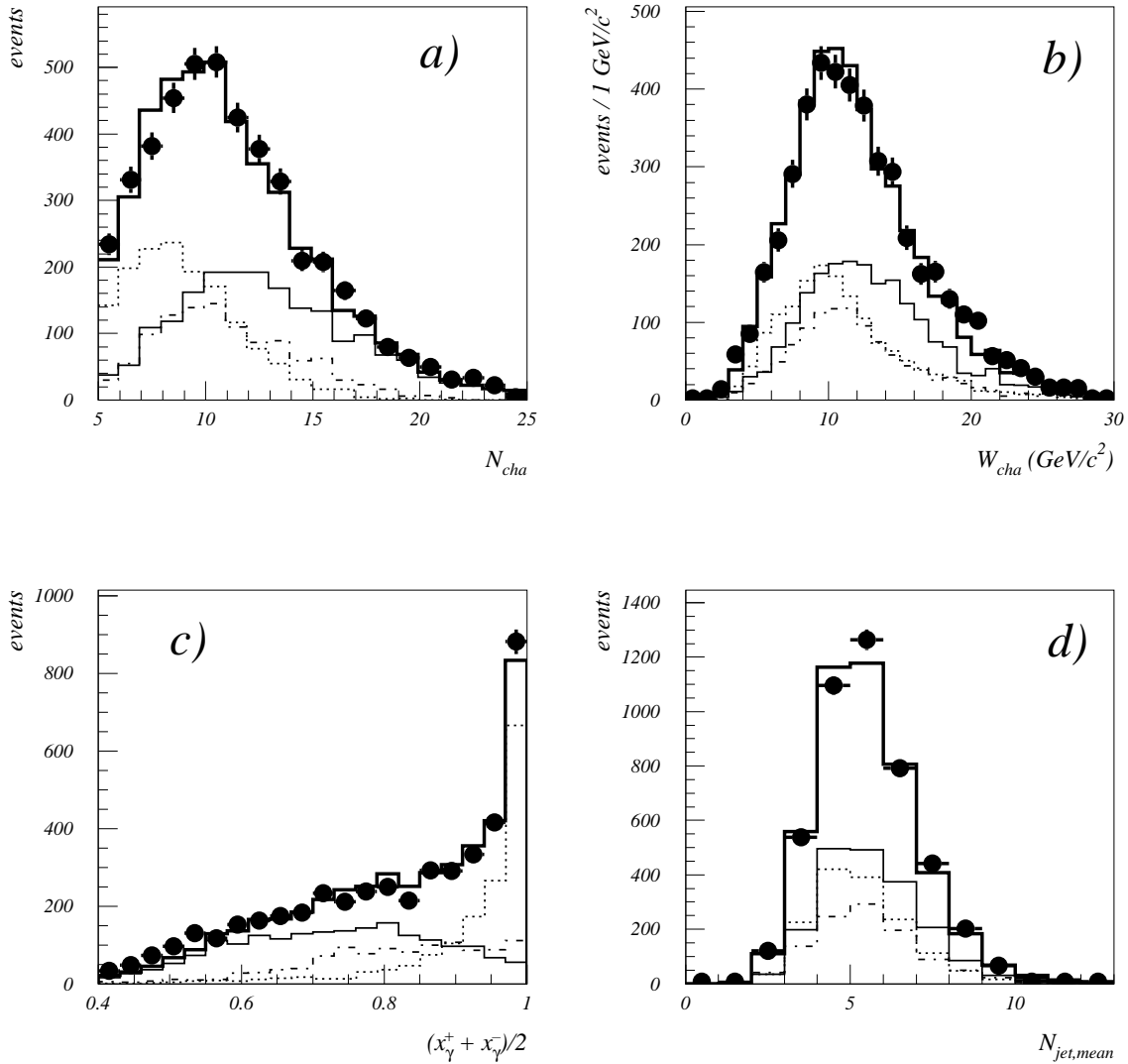,width=17cm}
\end{center}
\vspace*{-1cm}
\caption[]{Comparison of the background subtracted data with the simulation
after the fit for the charged multiplicity (a), the invariant mass
calculated with the charged particles only (b), ($x^+_\gamma+x^-_\gamma$)/2 (c) 
and the mean number of particles in the reconstructed jets (d). The simulation 
(thick solid histograms) sums all three model components scaled with the factors given in the text.
The contributions of \gaga subprocesses are shown by dotted (QPM term),
dashed-dotted (single-resolved) and thin solid (double-resolved) histograms.}
\label{fig:fig4}
\end{figure}

 Figure \ref{fig:fig4} presents the background-subtracted data compared to the 
\gaga simulation (with scaling factors applied) for the variables which were not involved in the fit.
The data are in good agreement with the simulation. The Monte Carlo set,
scaled according to the fit parameters, consists 
of 33\% QPM-like events, 23\% single-resolved and 44\% double-resolved events.
Double-resolved events are subdivided into contributions `no-MIA' (75\%) and `with-MIA' (25\%).
\subsection{Multiple parton interactions}
\par As already mentioned the modeling of the double-resolved events includes 
the so-called multiple parton
interactions (MIA) \cite{pyt}. Since the resolved photons are composite objects consisting 
of many partons, it is assumed that different pairwise interactions may take 
place during one \gaga collision. It has been shown that the inclusion of MIA
improves the description of the data \cite{h1}.
The amount of MIA in PYTHIA is determined by a cutoff parameter on the parton
transverse momentum. The default value of 1.6 GeV/c is used. MIA influence the model
predictions in two ways. The first is the case when the hard initiated jets satisfy 
the selection criteria and MIA systematically increase the jet transverse energy.
The second way is when the hard process alone does not provide 
two jets passing the selection 
criteria but combined with MIA initiated particles the event is selected.
The MIA contribution in this case is to the background and it has to be estimated 
and subtracted from the data according to the model expectation.
For this purpose PYTHIA is run in the mode with MIA switched off. It is found that
27\% of the double-resolved \gaga interactions with MIA contributions
would not pass the di-jet selection criteria if MIA were absent.
Note that the double-resolved part of the model is fitted in a coherent way, 
i.e. without splitting it into subsamples with-MIA and without it.
\par Summarizing the MIA influence, the model predicts the background coming
from MIA as 7\% (3\%) of the double-resolved part (the total model prediction).
On the other hand 18\% of the double-resolved events are affected by MIA via the increase of
the jet transverse momentum. The absolute value of the background coming
from the MIA contribution is estimated as $(131\pm7)$ events.
The importance of the 3-parameter fit performed has to be emphasized since 
the evaluation of the background is done by using the scale factors obtained.
\subsection{Hadronisation corrections, acceptance and background subtraction}
\par
\begin{figure}[ht!]
\begin{tabular}{cc}
\begin{minipage}[h]{9cm}
\hspace*{-1.cm}
\mbox{\epsfig{figure=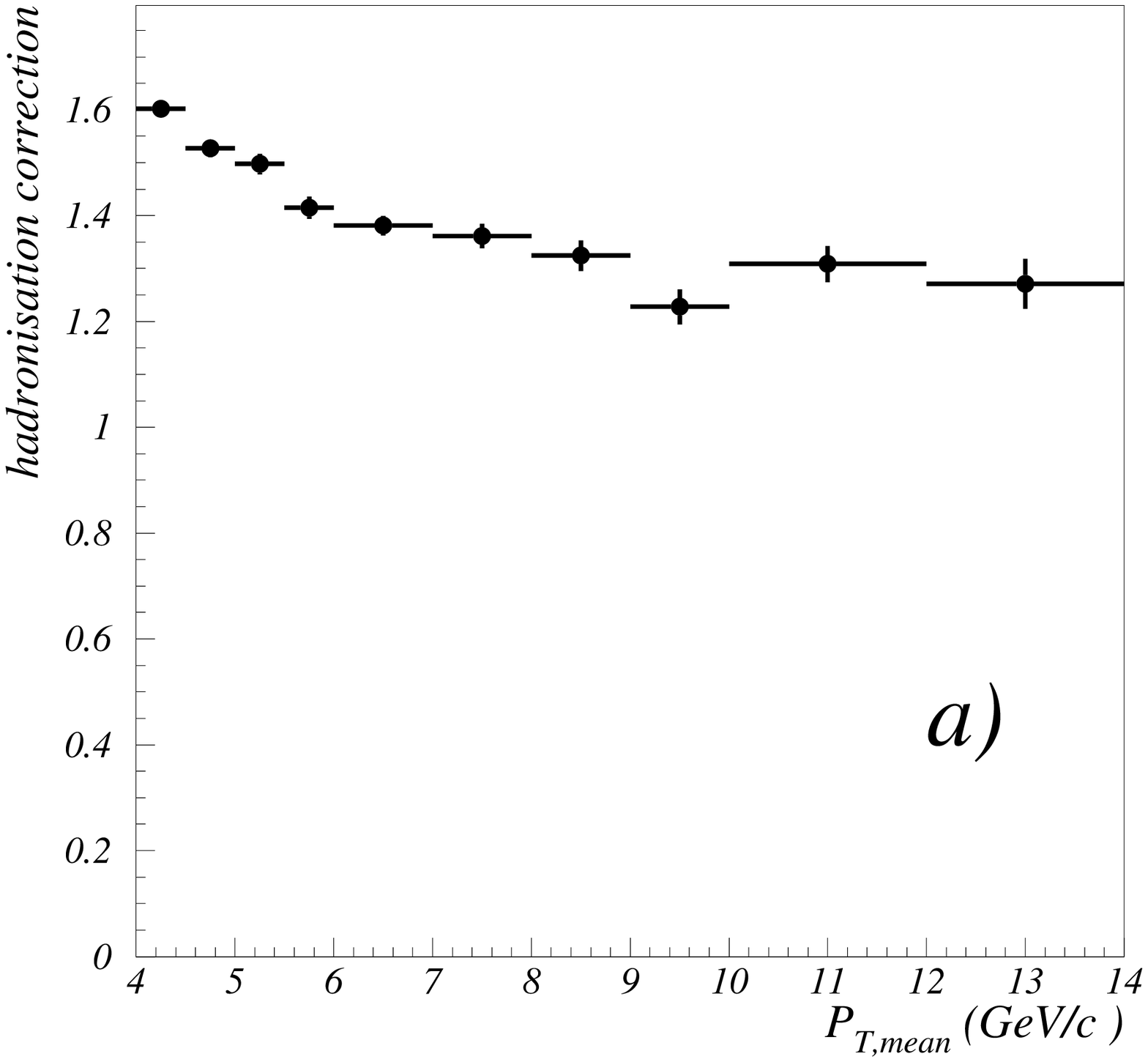,width=8.5cm}}
\end{minipage}
\begin{minipage}[h]{9cm}
\hspace*{-1.5cm}
\mbox{\epsfig{figure=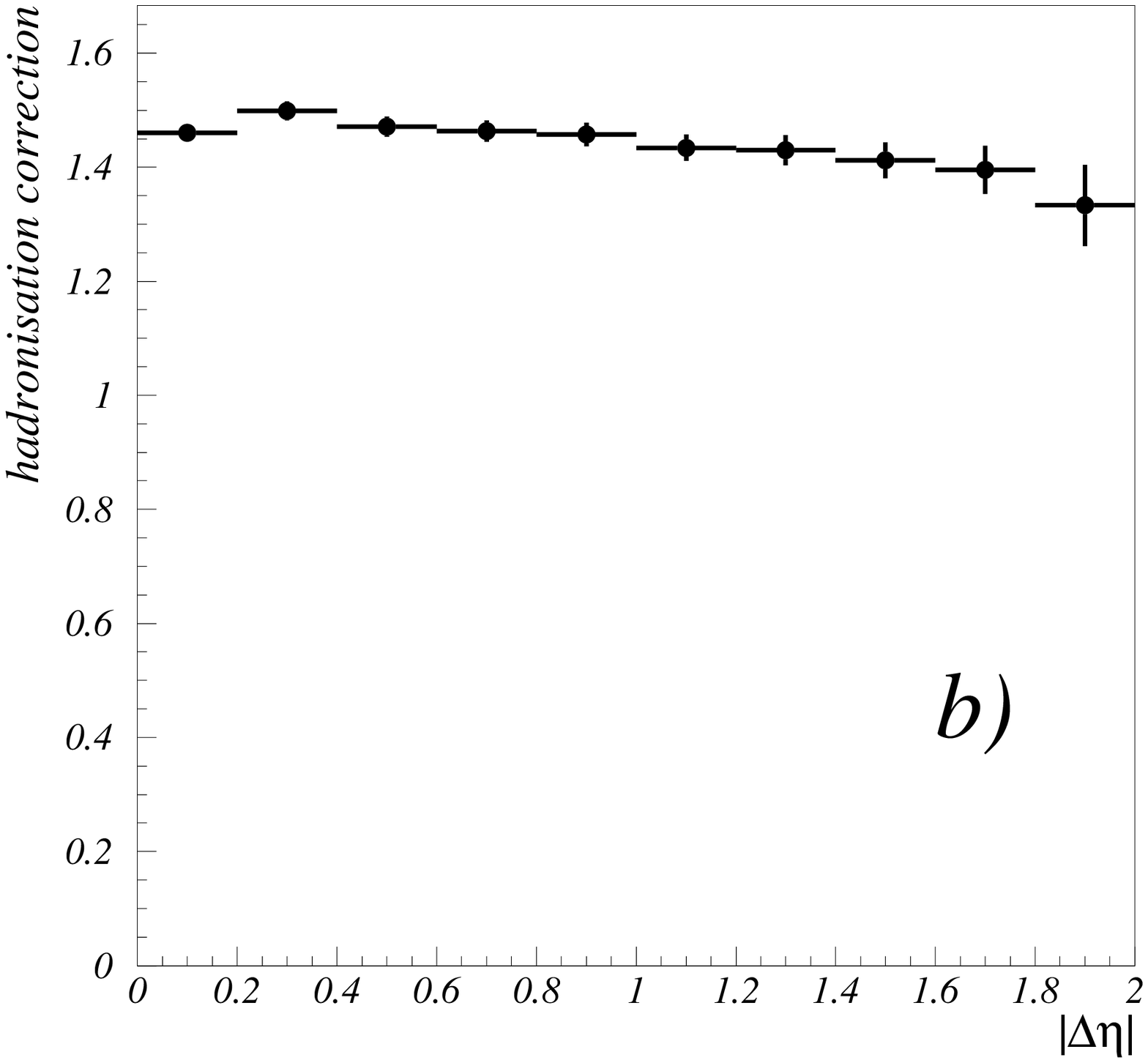,width=8.5cm}}
\end{minipage}
\end{tabular}
\caption[]{Hadronisation corrections as a function of the mean jet
transverse momentum \ptmean (a) and the absolute difference
of jet rapidities $|\Delta\eta|$ (b).}
\label{fig:fig5}
\end{figure}

 The theoretical predictions have to be transformed from the parton level 
to the level at which particles are produced in the $\gamma\gamma$ interaction vertex -
the particle level (`hadronisation corrections').
The Monte Carlo partons are considered in leading-order and are
not identical to the NLO partons in the theoretical calculations.
The corrections have been calculated using the PYTHIA program.
The distribution of the mean transverse momentum of the jets \ptmean
(difference of jet pseudo-rapidities $|\Delta\eta|$) obtained with the \kt 
at the parton level, is divided by the distribution obtained after
hadronisation. The resulting `hadronisation corrections' are shown
in Fig. \ref{fig:fig5}. They reach 60\% value at smallest $p_T$ and decrease to
30\% for higher transverse momentum.
\par
\begin{figure}[ht!]
\begin{tabular}{cc}
\begin{minipage}[h]{9cm}
\hspace*{-1.cm}
\mbox{\epsfig{figure=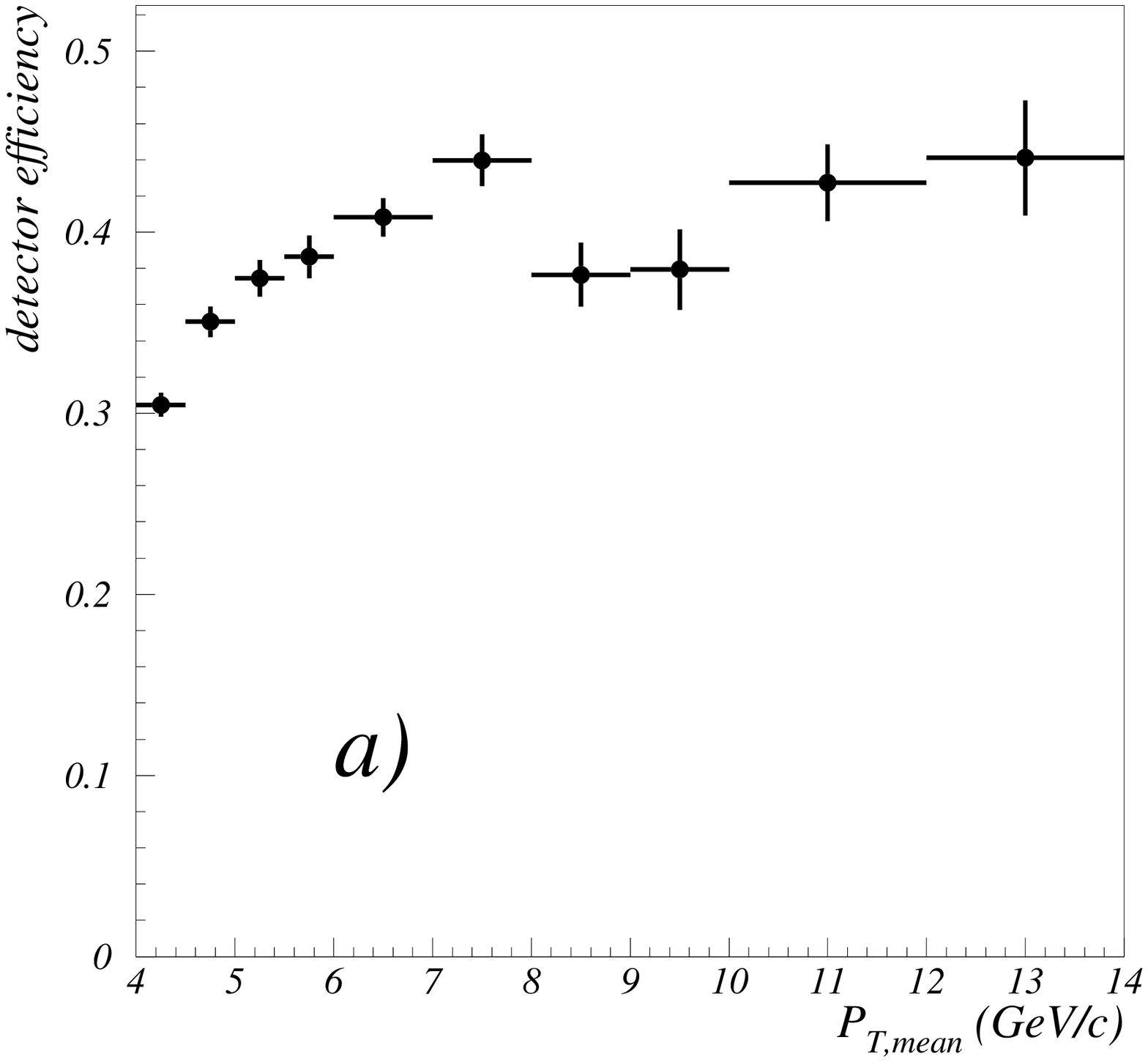,width=8.5cm}}
\end{minipage}
\begin{minipage}[h]{9cm}
\hspace*{-1.5cm}
\mbox{\epsfig{figure=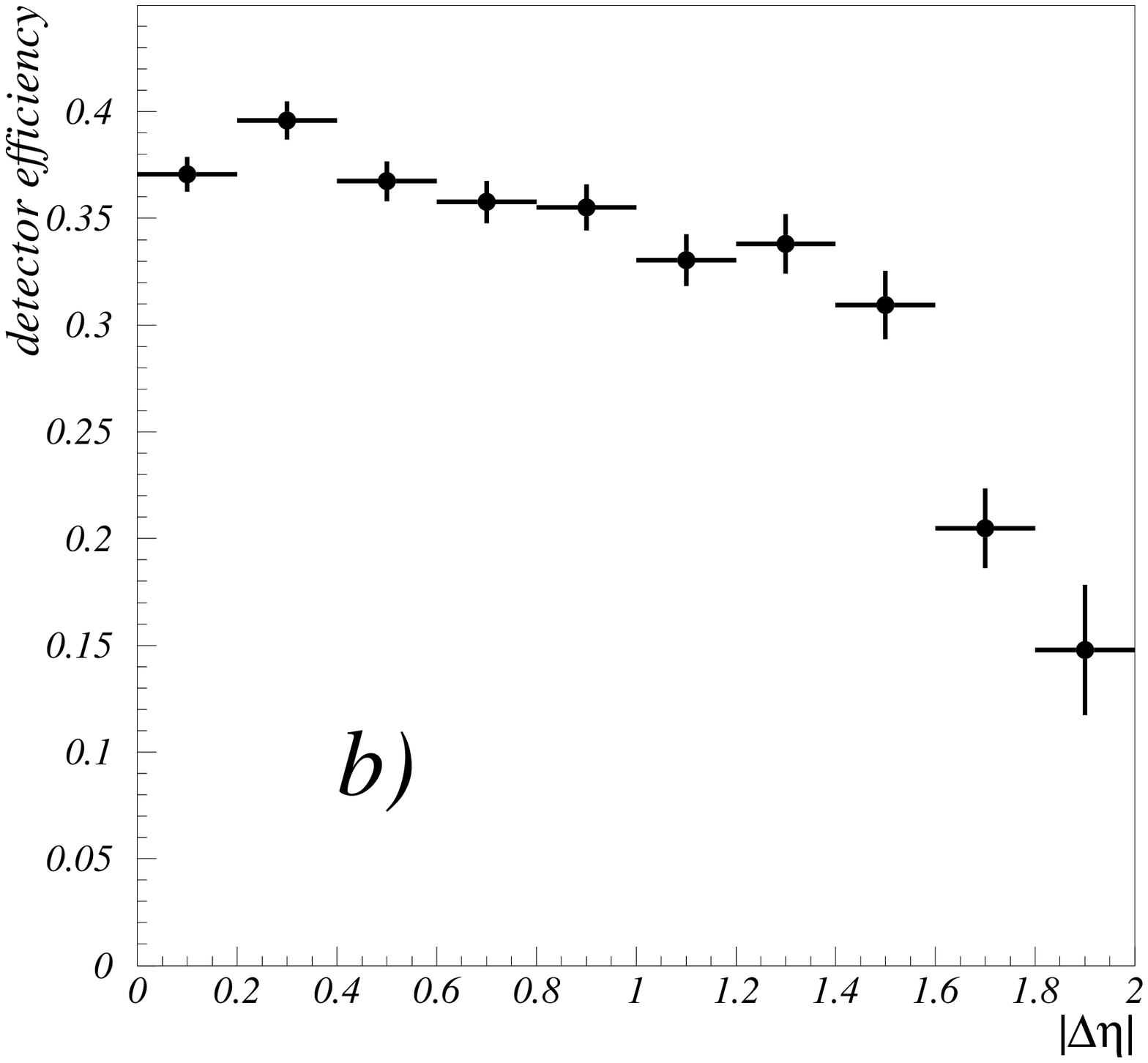,width=8.5cm}}
\end{minipage}
\end{tabular}
\caption[]{Detector efficiency as a function of \ptmean (a) and
$|\Delta\eta|$ (b).}
\label{fig:fig6}
\end{figure}

 The next step is to transform the data from the level of the detected 
particles to the hadrons produced in \agg interactions in order to
take the detector acceptance into account. The detector acceptance enters in two ways - 
either some events do not satisfy
the selection and thus they are lost or the observed distributions 
in the selected events are distorted. Bin-by-bin corrections are used.
Figure \ref{fig:fig6} illustrates the detector efficiency as a function
of \ptmean (the dip in the middle of the distribution is caused by limited statistics of
the simulation) and $|\Delta\eta|$.
\par Note that these corrections apply to two-jet events.
The first correction transfers the two-jet calculations carried out at the parton level to 
model expectations at the level of the produced particles.
The acceptance has been calculated both at the produced and detected particle levels
for the two-jet events.
That is why the following and last source of background 
(`non2-to-2 jets') should be taken into account. 
Some events are reconstructed with two jets because undetected particles change 
the event topology, but they do not have two jets at the production level.
This kind of background is estimated as $(893\pm13)$ events and it becomes the largest
source of background for the di-jet studies.

\section{Results and conclusions}

\par Figures 7(a-b) show the \ptmean and $|\Delta\eta|$ data distributions 
compared to the total simulation. All sources of the background mentioned above 
are given there as well.

\par The data are bin-to-bin corrected for the acceptance and, applying the integrated luminosity
of LEP corresponding to the data, the total di-jet cross-section is measured
to be $17.8\pm0.6$ pb for jets within the pseudo-rapidity 
range of $-1<\eta<1$ and for a jet transverse momentum $p_T$ above 3 GeV/c. The statistical and
systematic uncertainties (approximately of equal size) are added in quadrature. 
The systematic uncertainty includes the errors of the fit mentioned above, 
the uncertainty coming from MIA handling and the background estimations.
The model expectations are 
$20.5\pm0.1$ pb and $18.1\pm0.6$ pb for the calculations
carried out in the leading and next-to-leading order, respectively.
The differential cross-sections are obtained with bin-by-bin corrections for the detector 
efficiency (Fig.~6) for the data and with `hadronisation corrections' (Fig.~5)
applied to the theoretical calculations. The obtained cross-sections are shown 
in Fig. \ref{fig:fig8} and compared to the predictions mentioned 
above (the NLO calculations done with Monte Carlo method have sizable uncertainties which 
are shown on the plots). Numerical results are presented in Table \ref{tab:tab1} 
as a function of \ptmean and in Table \ref{tab:tab2} as a function of $|\Delta\eta|$.
The errors include statistical and systematic uncertainties added in quadrature. 
\par Since we scaled the contribution of QPM-like events (`direct' term) and this
part is well calculable, it seems interesting to compare the scaled distributions
to the results of calculations carried out in the leading and next-to-leading
orders. The results are presented in Fig. \ref{fig:fig9}. The scaled `direct' term 
is only around 10\% below NLO calculations; the main difference comes
from the small \ptmean domain, while there is good agreement with NLO calculations
for \ptmean greater than 4.5 GeV/c.

\par In conclusion, the production of two high-$p_T$ jets in the interactions 
of quasi-real photons is studied with the DELPHI data taken at LEP-2 with an 
integrated $e^+e^-$ luminosity of 550 pb$^{-1}$. The jets reconstructed by the \kt 
are defined within the pseudo-rapidity range of $-1<\eta<1$  
and for a jet transverse momentum $p_T$ above 3 GeV/c. The total and
differential di-jet cross-sections are measured for a mean jet momentum 
\ptmean from 4 GeV/c to 14 GeV/c. The total cross-section agrees with
the next-to-leading order perturbative QCD calculation within the experimental
uncertainties and is 18$\%$ below the calculation carried out
in the leading order. The measured differential di-jet cross-section is
also found to be in good agreement with NLO QCD predictions \cite{theor} and
complements results obtained by other LEP experiments \cite{otheropal,otherl3}.

\subsection*{Acknowledgements}
\vskip 3 mm
We are greatly indebted to our technical 
collaborators, to the members of the CERN-SL Division for the excellent 
performance of the LEP collider, and to the funding agencies for their
support in building and operating the DELPHI detector.\\
We acknowledge in particular the support of \\
Austrian Federal Ministry of Education, Science and Culture,
GZ 616.364/2-III/2a/98, \\
FNRS--FWO, Flanders Institute to encourage scientific and technological 
research in the industry (IWT) and Belgian Federal Office for Scientific, 
Technical and Cultural affairs (OSTC), Belgium, \\
FINEP, CNPq, CAPES, FUJB and FAPERJ, Brazil, \\
Ministry of Education of the Czech Republic, project LC527, \\
Academy of Sciences of the Czech Republic, project AV0Z10100502, \\
Commission of the European Communities (DG XII), \\
Direction des Sciences de la Mati$\grave{\mbox{\rm e}}$re, CEA, France, \\
Bundesministerium f$\ddot{\mbox{\rm u}}$r Bildung, Wissenschaft, Forschung 
und Technologie, Germany,\\
General Secretariat for Research and Technology, Greece, \\
National Science Foundation (NWO) and Foundation for Research on Matter (FOM),
The Netherlands, \\
Norwegian Research Council,  \\
State Committee for Scientific Research, Poland, SPUB-M/CERN/PO3/DZ296/2000,
SPUB-M/CERN/PO3/DZ297/2000, 2P03B 104 19 and 2P03B 69 23(2002-2004),\\
FCT - Funda\c{c}\~ao para a Ci\^encia e Tecnologia, Portugal, \\
Vedecka grantova agentura MS SR, Slovakia, Nr. 95/5195/134, \\
Ministry of Science and Technology of the Republic of Slovenia, \\
CICYT, Spain, AEN99-0950 and AEN99-0761,  \\
The Swedish Research Council,      \\
The Science and Technology Facilities Council, UK, \\
Department of Energy, USA, DE-FG02-01ER41155, \\
EEC RTN contract HPRN-CT-00292-2002. \\


\clearpage

\begin{center}
\begin{table}
\begin{tabular}{|c|c|c|c|} \hline
$p_{T,mean}$ & $\sigma_{measured, total}$ (pb/GeV) & $\sigma_{LO}$ (pb/GeV) & $\sigma_{NLO}$ 
(pb/GeV) \\ \hline
4.0-4.5 & $9.59\pm0.53$ & $11.59\pm0.13$ & $9.75\pm0.85$ \\ \hline
4.5-5.0 & $6.71\pm0.41$ & $7.61\pm0.10$ & $6.71\pm0.53$ \\ \hline
5.0-5.5 & $4.56\pm0.33$ & $5.24\pm0.07$ & $4.49\pm0.27$ \\ \hline
5.5-6.0 & $3.06\pm0.26$ & $3.86\pm0.06$ & $3.49\pm0.19$ \\ \hline
6.0-7.0 & $2.19\pm0.16$ & $2.46\pm0.04$ & $2.28\pm0.12$ \\ \hline
7.0-8.0 & $1.17\pm0.12$ & $1.40\pm0.03$ & $1.28\pm0.08$ \\ \hline
8.0-9.0 & $0.88\pm0.12$ & $0.88\pm0.02$ & $0.79\pm0.06$ \\ \hline
9.0-10.0 & $0.71\pm0.11$ & $0.61\pm0.02$ & $0.58\pm0.05$ \\ \hline
10.0-12.0 & $0.31\pm0.05$ & $0.33\pm0.01$ & $0.31\pm0.03$ \\ \hline
12.0-14.0 & $0.11\pm0.03$ & $0.18\pm0.01$ & $0.16\pm0.03$ \\ \hline
\end{tabular}
\caption[]{Measured cross-section of di-jet production in quasi-real 
\agg interactions as a function of \ptmean together with
leading order and next-to-leading order calculations \cite{theor}.}
\label{tab:tab1}
\end{table}

\begin{table}
\begin{tabular}{|c|c|c|c|} \hline
$|\Delta\eta|$ & $\sigma_{measured, total}$ (pb) & $\sigma_{LO}$ (pb) & $\sigma_{NLO}$ (pb) \\ \hline
0.-0.2 & $3.50\pm0.21$ & $4.23\pm0.05$ & $3.44\pm0.25$ \\ \hline
0.2-0.4 & $2.96\pm0.18$ & $3.66\pm0.05$ & $2.76\pm0.20$ \\ \hline
0.4-0.6 & $2.54\pm0.18$ & $3.31\pm0.04$ & $3.32\pm0.25$ \\ \hline
0.6-0.8 & $2.67\pm0.19$ & $2.81\pm0.04$ & $2.36\pm0.19$ \\ \hline
0.8-1.0 & $2.09\pm0.17$ & $2.35\pm0.04$ & $2.20\pm0.24$ \\ \hline
1.0-1.2 & $1.78\pm0.16$ & $1.87\pm0.03$ & $1.84\pm0.20$ \\ \hline
1.2-1.4 & $1.39\pm0.14$ & $1.38\pm0.03$ & $1.28\pm0.13$ \\ \hline
1.4-1.6 & $0.67\pm0.11$ & $0.95\pm0.02$ & $0.80\pm0.09$ \\ \hline
1.6-1.8 & $0.29\pm0.12$ & $0.54\pm0.02$ & $0.63\pm0.09$ \\ \hline
1.8-2.0 & $0.29\pm0.11$ & $0.16\pm0.02$ & $0.09\pm0.04$ \\ \hline
\end{tabular}
\caption[]{Measured cross-section of di-jet production in quasi-real 
\agg interactions as a function of $|\Delta\eta|$ together with
leading order and next-to-leading order calculations \cite{theor}.}
\label{tab:tab2}
\end{table}
\end{center}

\clearpage

\vspace*{1cm}
\begin{figure}[ht!]
\begin{tabular}{cc}
\begin{minipage}[h]{9cm}
\hspace*{-1.cm}
\mbox{\epsfig{figure=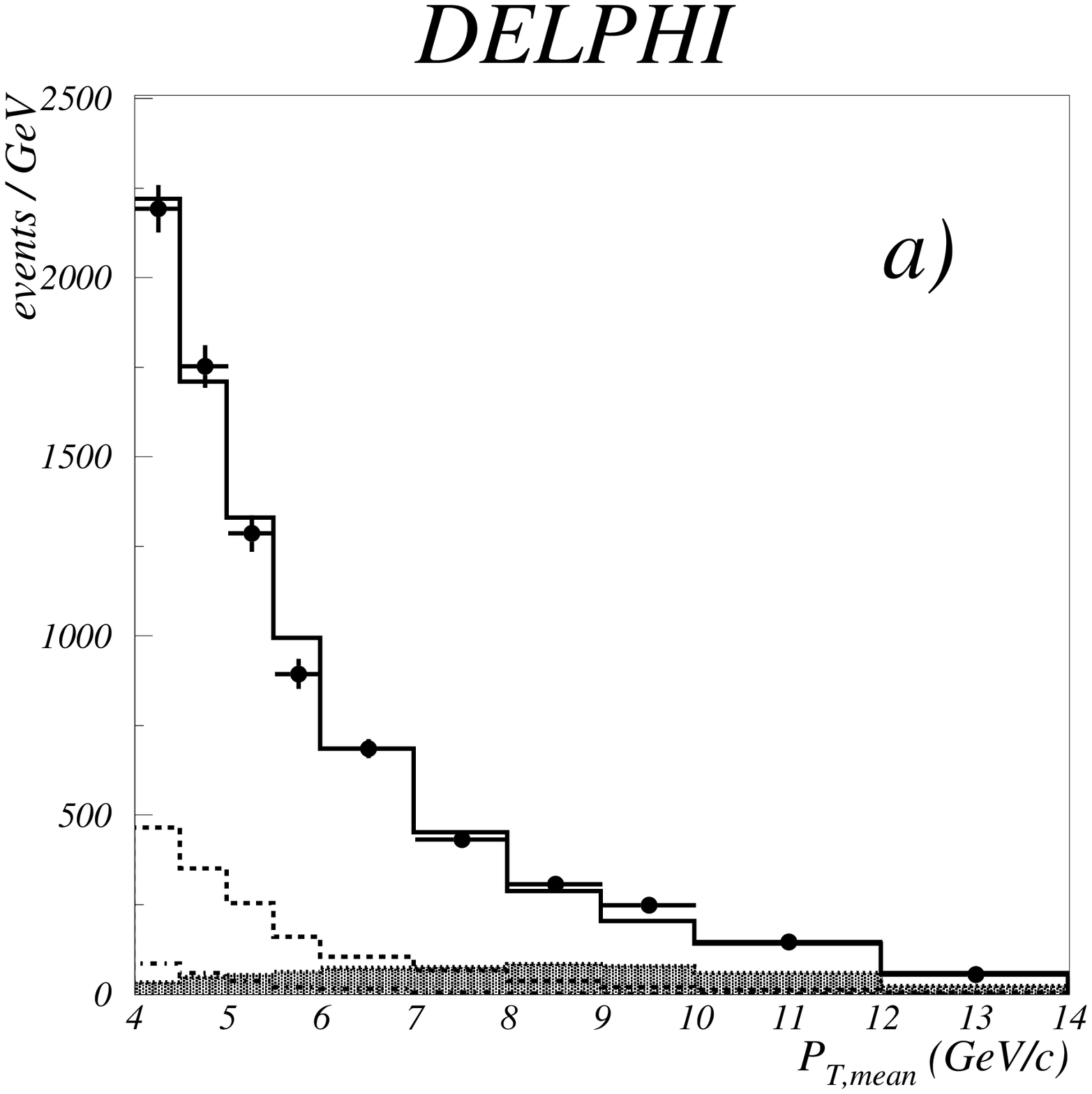,width=8.5cm}}
\end{minipage}
\begin{minipage}[h]{9cm}
\hspace*{-1.5cm}
\mbox{\epsfig{figure=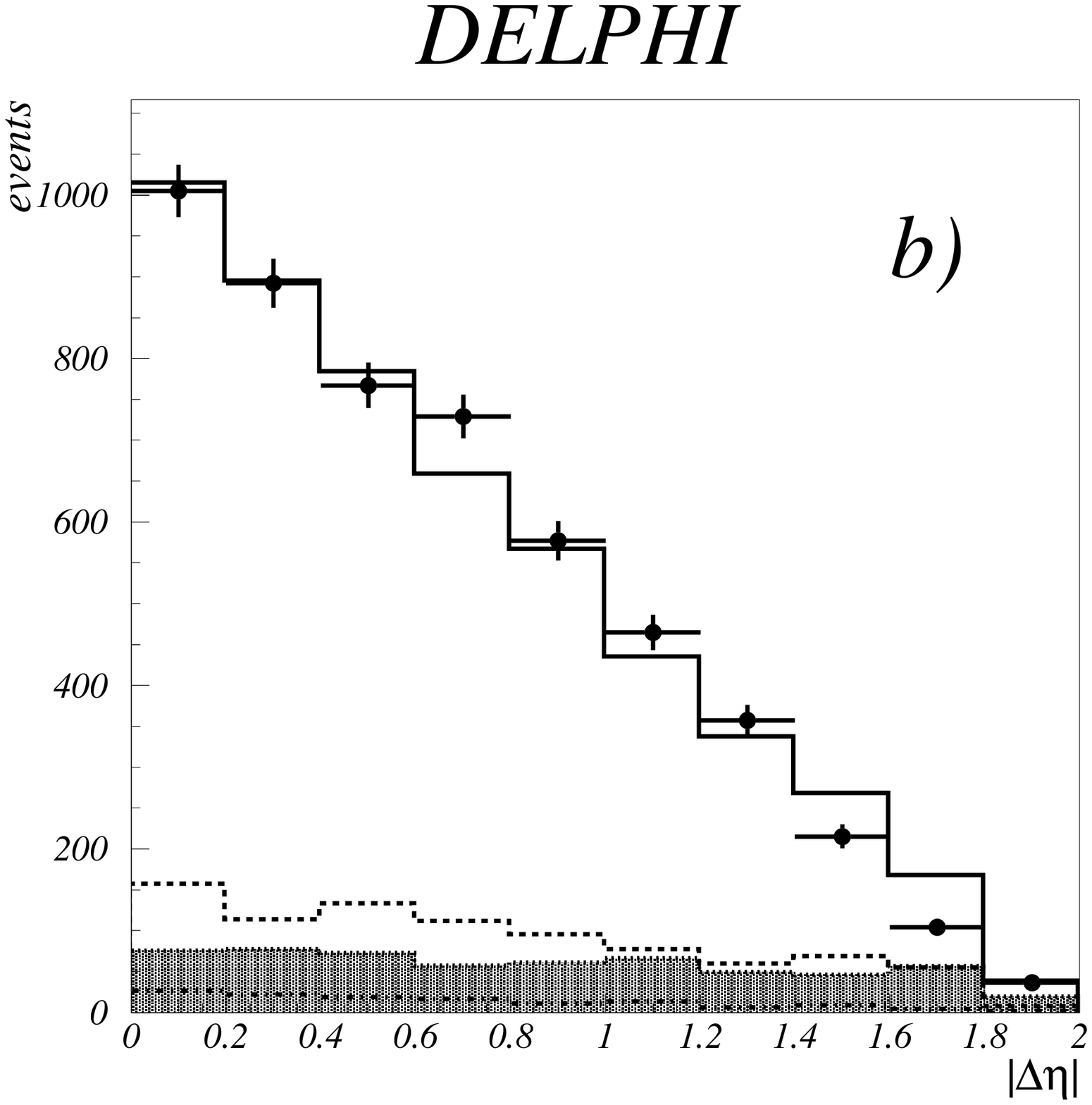,width=8.5cm}}
\end{minipage}
\end{tabular}
\caption[]{Distributions of the mean jet transverse momentum \ptmean (a) 
and the absolute difference of jet rapidities $|\Delta\eta|$ (b).
The data are presented by bars, the sum of the scaled PYTHIA prediction and
of the background by solid histograms. The background components
are shown by dash-dotted lines (MIA contribution), by dashed lines (`non2-to-2 jets' background) 
and by shaded histograms (`non-$\gamma\gamma$' background).}
\label{fig:fig7}
\end{figure}

\clearpage
\begin{figure}[ht!]
\begin{minipage}[h]{11cm}
\hspace*{1.cm}
\mbox{\epsfig{figure=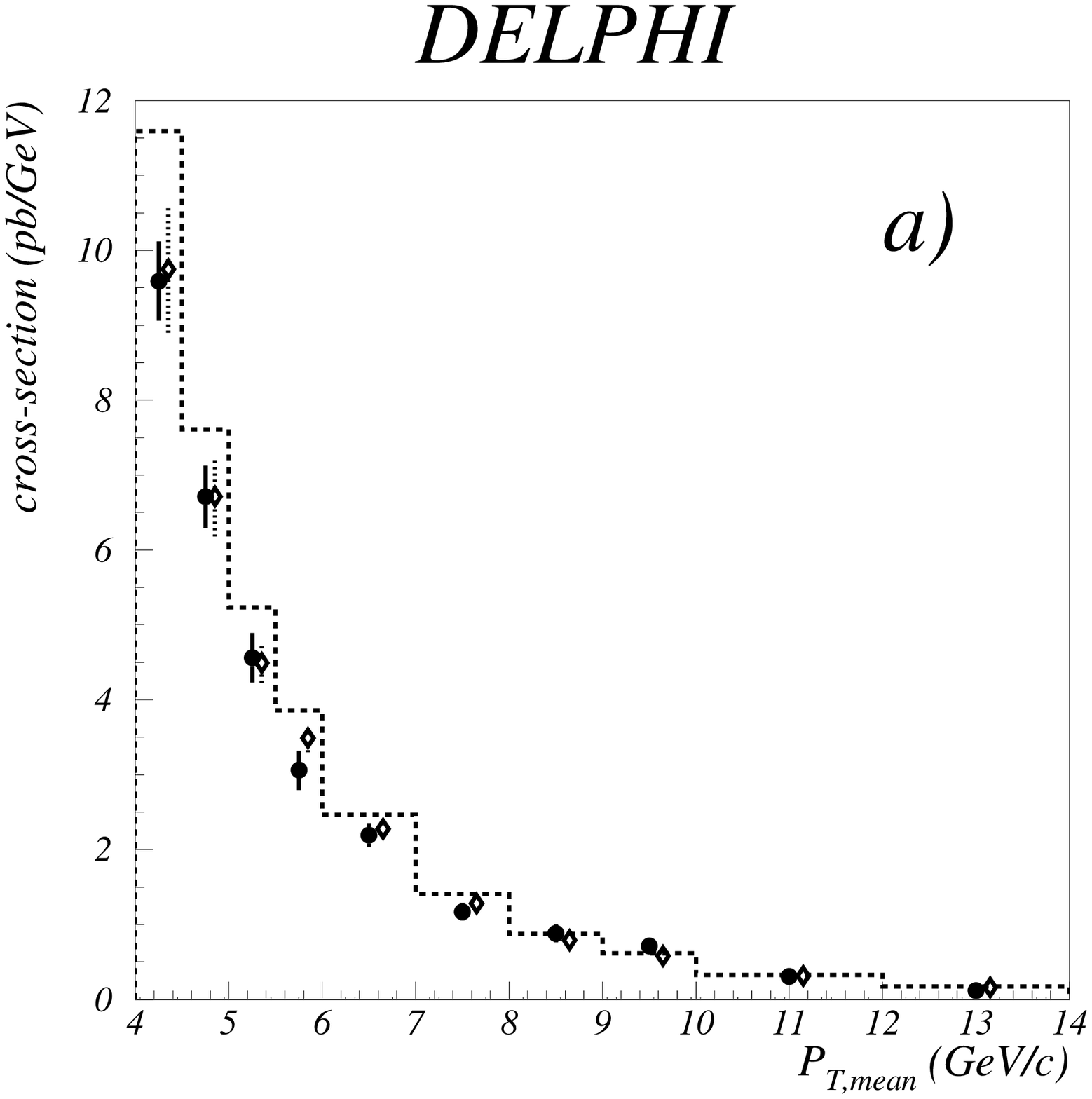,width=10cm}}
\end{minipage} \\
\begin{minipage}[h]{11cm}
\vspace*{1cm}
\hspace*{1.5cm}
\mbox{\epsfig{figure=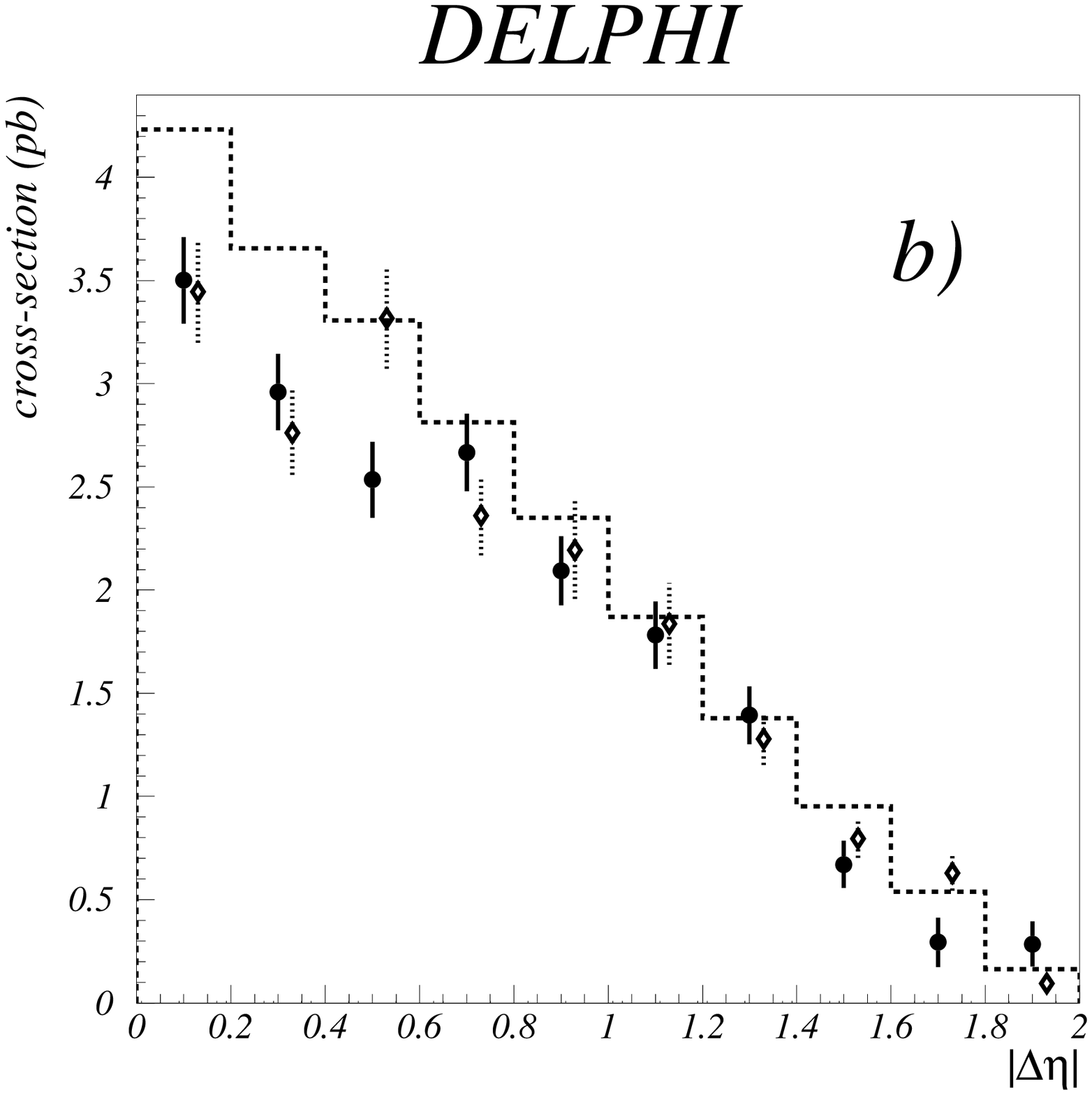,width=10cm}}
\end{minipage}
\caption[]{Cross-section of di-jet production in quasi-real 
\agg interactions as a function of \ptmean (a) and $|\Delta\eta|$ (b). 
The jets are reconstructed by the \kt within the
pseudo-rapidity range of $-1<\eta<1$  and the jet transverse
momentum $p_T$ above 3 GeV/c. Dashed histograms show the leading order calculations and
next-to-leading order calculations \cite{theor} are presented by open diamonds.}
\label{fig:fig8}
\end{figure}
\begin{figure}[ht!]
\begin{tabular}{cc}
\begin{minipage}[h]{9cm}
\hspace*{-1.cm}
\mbox{\epsfig{figure=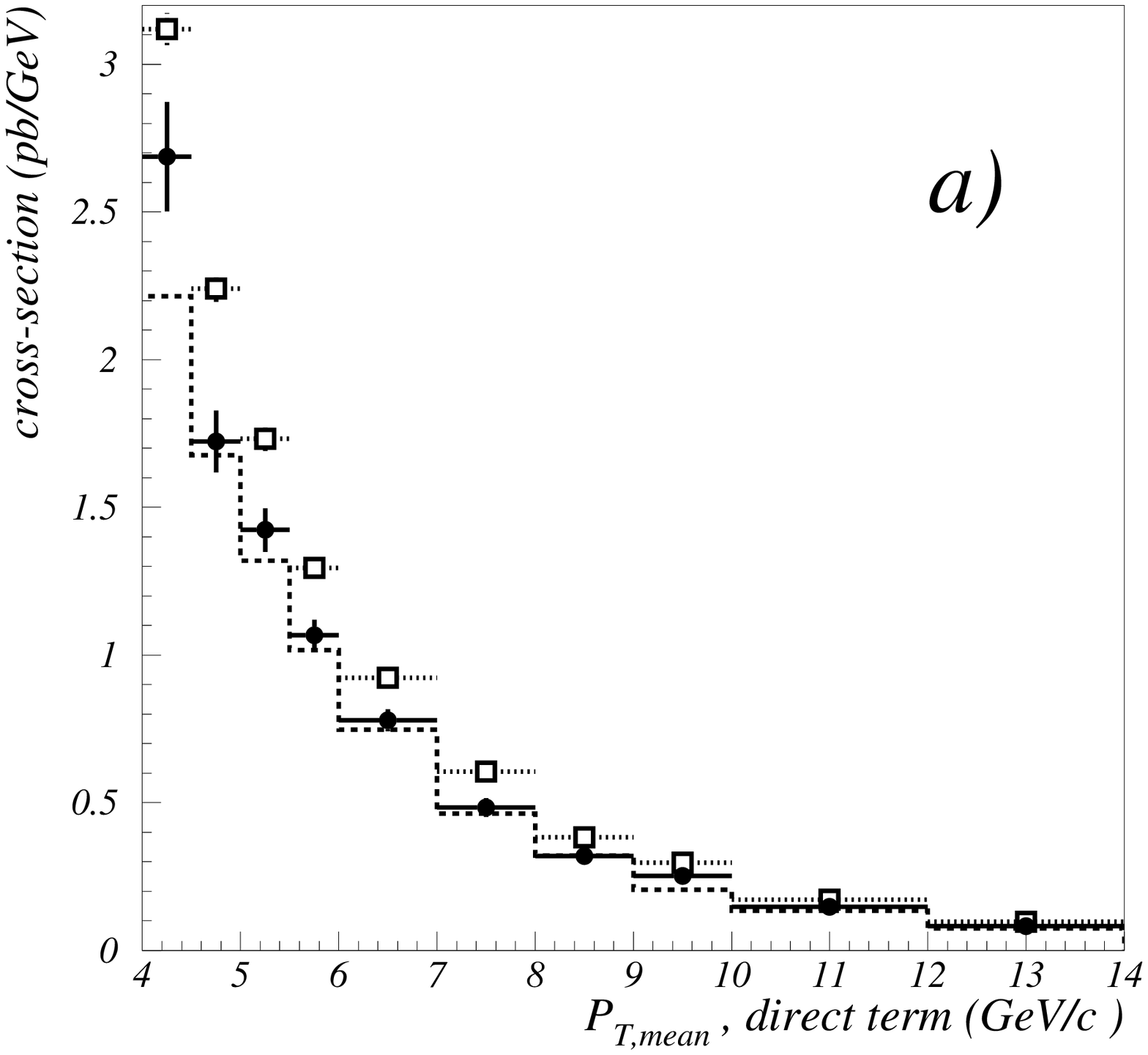,width=8.5cm}}
\end{minipage}
\begin{minipage}[h]{9cm}
\hspace*{-1.5cm}
\mbox{\epsfig{figure=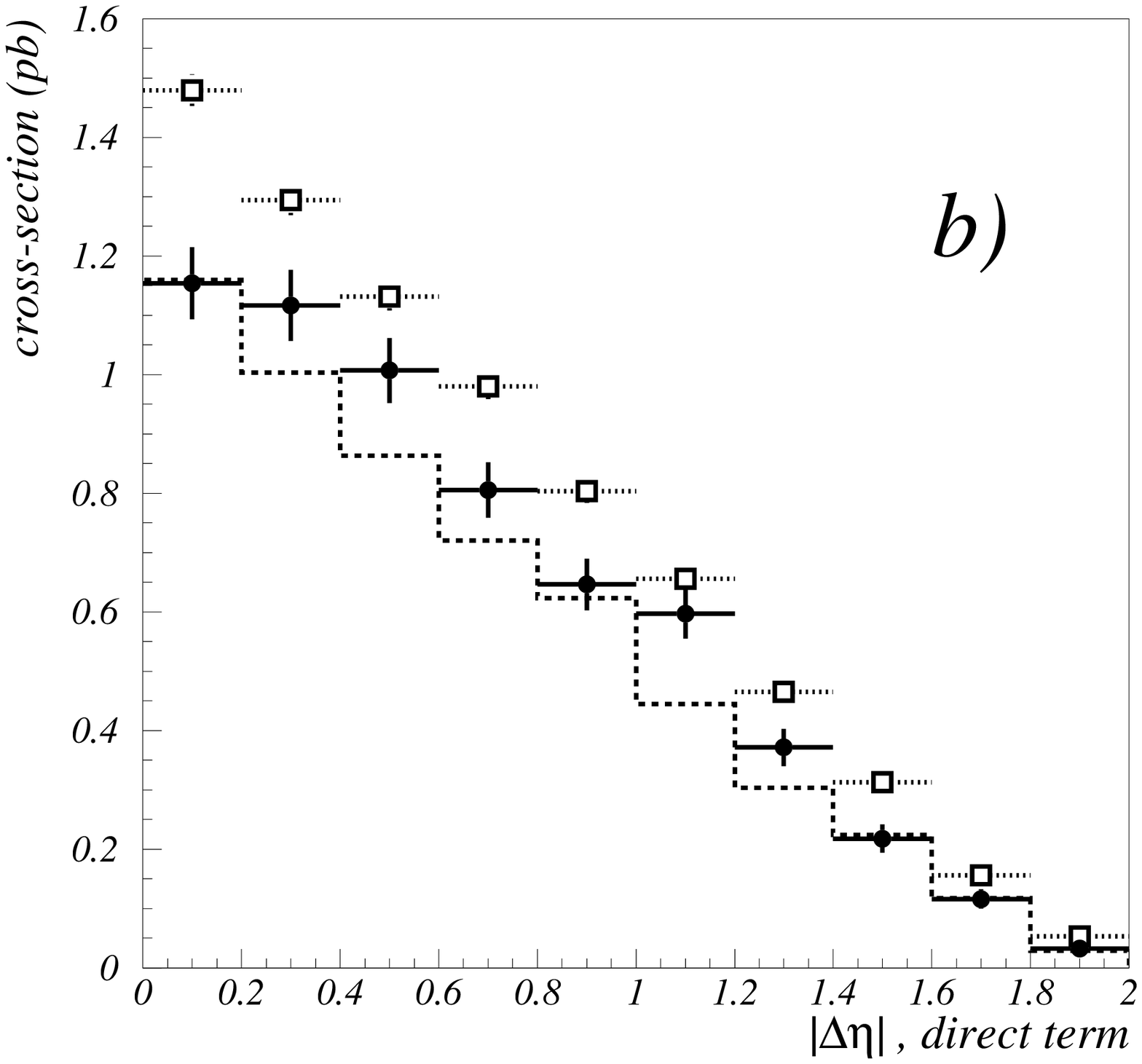,width=8.5cm}}
\end{minipage}
\end{tabular}
\caption[]{Cross-section of the `direct' term as a function of \ptmean (a) and $|\Delta\eta|$ (b).
Dashed histograms correspond to the PYTHIA predictions scaled according to the fit.
The circles (squares) present the results of the NLO (LO) calculations \cite{theor}.}
\label{fig:fig9}
\end{figure}

\end{document}